\documentclass[3p,twocolumn,sort&compress,showpacs,showkeys,preprintnumbers,amsmath,amssymb,aps,floatfix,pre]{elsarticle}
 
 \usepackage{lineno}
\modulolinenumbers[5]

\usepackage{graphicx}
\usepackage{epstopdf}
\usepackage{amsmath}
\usepackage{multirow}
\usepackage{times}
\usepackage{amssymb}
\usepackage{dcolumn}%
\usepackage{bm}%
\usepackage[dvipdfmx]{hyperref}%
\usepackage{upgreek}
\usepackage{wasysym}
\usepackage{verbatim}

\hypersetup
{
	colorlinks=true,
	linkcolor=blue,
	citecolor=blue
}

\journal{Journal of \LaTeX\ Templates}

\begin{document}

\begin{frontmatter}

\title{Elastic modulus and yield strength of semicrystalline polymers with bond disorder are higher than in atomic crystals}
\author{A. Giuntoli}
\address{Dipartimento di Fisica ``Enrico Fermi'', 
Universit\`a di Pisa, Largo B.\@Pontecorvo 3, I-56127 Pisa, Italy}


\author{D. Leporini*}
\address{Dipartimento di Fisica ``Enrico Fermi'', 
Universit\`a di Pisa, Largo B.\@Pontecorvo 3, I-56127 Pisa, Italy}
\address{IPCF-CNR, UOS Pisa, Italy}

\cortext[D. Leporini]{Corresponding author}
\ead{dino.leporini@unipi.it}

\begin{abstract}
We perform thorough molecular-dynamics simulations to compare elasticity and yielding of atomic crystals and model semicrystalline polymers,
the latter characterized by very similar {\it positional ordering} with respect to atomic crystals and considerable {\it bond disorder}. We find that the elastic modulus $G$, the shear yield strength, $\tau_Y$, 
and the critical yield strain $\epsilon_c$ of semicrystalline polymers are higher {than ( $G$, $\tau_Y$)}, or comparable {to ($\epsilon_c$)}, the corresponding ones of atomic crystals. 
The findings suggest that the bond disorder suppresses dislocation-mediated plasticity 
in polymeric solids with positional order.
\end{abstract}

\end{frontmatter}

\noindent keywords: Molecular-dynamics simulations, elasticity, plasticity, semicrystalline polymers

\section{Introduction}
Elasticity theories \cite{ArgonPlasticTheory,EshelbyElastic,BudianskyElastic,WardElastic} predict that solid materials respond linearly with elastic modulus $G$ to small shear deformations. 
Upon increasing strain, amorphous solids show complex and far from linear behavior \cite{BarratPlastic02,BarratPlastic04,BarratPlastic05}.
When a critical yield strain $\epsilon_c$ is reached, corresponding to the shear yield strength $\tau_Y$, the transition from the (reversible) elastic state to the (irreversible) plastic 
state takes place \cite{StachurskiProgPolymSci97,RobbinsHoyJPolymSci06,RottlerSoftMatter10}. In an ideal elasto-plastic body (Hooke-St.Venant) $\tau_Y$ 
is the maximum stress \cite{StachurskiProgPolymSci97}.

It is well-known that plasticity in crystalline solids results from the structure and the mobility of defects (in particular dislocations)  \cite{HirthDislocationBook}.  Dislocations do not exist in amorphous
polymers, but, under an applied stress, elementary shear displacements can occur in a spatially correlated linear {domain} which can close on itself to form a loop to be interpreted in terms 
of classical dislocation
mechanics and energetics \cite{GilmanPlastic68,Li_dislocation,StachurskiProgPolymSci97,BowdenSahaDislocationPolymYield74}. However, even if the model can be used to fit the experimental data, there are 
conceptual problems to extend dislocation based concepts to glassy polymers  \cite{ArgonFractures,LamChongPlasticityPolym99}. That difficulty is part of the complexities involved in the  phenomenon of plastic
deformation in glassy polymers which is not yet fully understood, in spite of many accurate phenomenological models, see e.g. refs. \cite{ArgonFractures,StachurskiProgPolymSci97} for comprehensive reviews. 
In particular, Argon considered a scenario where individual chains are embedded in an elastic continuum \cite{ArgonFractures}. He argued that plastic deformation is caused by the cooperative rearrangements 
of a  cluster of segments with size $\Omega_f$. The latter region is thermally activated under the applied stress to overcome the resistance that is generated from elastic interaction of the polymer chain 
with its surroundings. $\Omega_f$ is significantly smaller than the activation volume of dislocations  \cite{ArgonFractures,StachurskiProgPolymSci97}. The concept of localized cooperative rearrangements was 
proven to be fruitful also to account for the plasticity of non-polymeric glasses \cite{ArgonFractures}. It was found that $\Omega_f$ is much smaller in amorphous metals with respect to glassy polymers. 
In comparison with the plasticity of crystalline solids, where the long-range positional order permits the translation of dislocations, the plasticity of disordered solids is mainly driven by the 
activation of cooperative rearrangements within the cluster of segments  \cite{ArgonFractures,HoEtAlMM03}.

The previous discussion highlights that there are strong differences in the microscopic mechanisms of plasticity of {\it atomic crystals} and {\it polymeric glasses}. These two classes of  materials 
differ in two rather distinct aspects, namely the connectivity and the positional ordering. Since these two features cannot be thought of as mutually independent  and may exhibit antagonism, singling out 
the role of each of them is of interest. As a first step along this direction, the present paper aims at elucidating the role of connectivity into the linear and non-linear deformation of solids with {\it different
connectivity} and {\it rather similar positional order}. Influence of connectivity outside the elastic limit has been recently reviewed  \cite{GreavesNatMat2011}.
Our study considers atomic crystals and polymer semicrystals, the latter with very similar positional ordering and considerable bond disorder to average out the coupling between connectivity and
positional order \cite{GiuntoliCristallo,NicolaElastico}. 
We find that the elastic modulus $G$, the shear yield strength, $\tau_Y$, and the critical yield strain $\epsilon_c$ of polymeric semicrystals are higher {than ( $G$, $\tau_Y$)}, 
or comparable {to ($\epsilon_c$)},
the corresponding ones of atomic crystals. The results show that the introduction of disordered connectivity perturbs the long-range order, most presumably suppressing dislocation-mediated plasticity, and 
then increases the shear strength. In this sense,  {\it if positional order is present}, atomic and polymeric plasticity appear to be {not reconcilable}.
It is worth noting that that the previous conclusion does not hold for glassy systems where, e.g., the plasticity of polymeric and atomic glasses with different connectivity exhibits similarities
\cite{ArgonSilicon}.

\section{Methods}
\label{methods}
Molecular-dynamics (MD) numerical simulations were carried out on two different systems, i.e. a melt of linear polymers and an atomic liquid.

As to the polymer systems, a coarse-grained polymer model of $N_c=50$ linear, fully-flexible, unentangled chains with $M=10$ monomers per chain is considered \cite{GiuntoliCristallo}.
The total number of monomers is $N=500$. 
Non-bonded monomers at distance $r$ belonging to the same or different chain interact via the truncated Lennard-Jones (LJ) potential: 
\begin{equation}
\label{eq1}
U^{LJ}(r)=\varepsilon\left [ \left (\frac{\sigma^*}{r}\right)^{12 } - 2\left (\frac{\sigma^*}{r}\right)^6 \right]+U_{cut}
\end{equation}
$\sigma^*=2^{1/6}\sigma$ is the position of the potential minimum with depth $\varepsilon$. The value of the constant
$U_{cut}$ is chosen to ensure $U^{LJ}(r)=0$ at $r \geq r_c=2.5\,\sigma$. The bonded monomers interaction is described by an harmonic potential $U^b$:
\begin{equation}
 U^b(r)=k(r-r_0)^2
\end{equation}
The parameters $k$ and $r_0$
have been set to $2500 \, \varepsilon  / \sigma^2 $ and $ 0.97\,\sigma $ respectively \cite{GrestPRA33}. Full-flexibility of the chain is ensured by the missing bending stiffness between 
adjacent bonds \cite{NicolaElastico}. {It must be pointed out that the bond length $\simeq 0.97\,\sigma $ prevents the significant heterogeneity of the monomer arrangements which is seen with longer bond length, see Fig.6a 
of ref. \cite{NicolaElastico}.}

As to the atomic systems we consider systems of $N=500$ atoms interacting with the truncated Lennard-Jones potential as in Eq. \ref{eq1}.

From this point on, all quantities are expressed in term of reduced units: 
lengths in units of $\sigma$, temperatures in units 
of $\varepsilon/k_B$ (with $k_B$ the Boltzmann constant) and time $t_{MD}$ in units of $\sigma \sqrt{m / \varepsilon}$ where 
$m$ is the monomer mass. We set $m = k_B = 1$. Periodic boundary conditions are used. The study was performed
in the $NPT$ ensemble (constant number of particles, pressure and temperature). The integration time step is set to $\Delta t=0.003$ time units \cite{Puosi11,UnivPhilMag11,PuosiLepoJCPCor12,UnivSoftMatter11}
The simulations were carried out using LAMMPS molecular dynamics software (http://lammps.sandia.gov) \cite{PlimptonLAMMPS}.

Fifty-six polymeric samples with initial different random monomer positions and velocities are equilibrated at temperature $T=0.7$ and pressure $P=4.7$, corresponding to number density $\rho \sim 1$. 
{That thermodynamic states allows the polymer melt to equilibrate in the liquid phase for at least three times the average reorientation time of the end-end vector of the chain.}
After the equilibration, production runs started and proceeded up to the spontaneous onset and the full development of the crystallization of the samples. Fourteen runs failed to crystallize in a 
reasonable amount of time, while forty-two of them underwent crystallization forming polymorph crystals with distorted body-centered cubic (Bcc) lattices. 
Additional details, in particular concerning the crystallization process, are given elsewhere \cite{GiuntoliCristallo}.
Sixty-four atomic liquid runs were equilibrated with starting temperature $T=1.5$ and pressure $P=20.0$. The temperature is higher in the atomic systems to avoid  crystallization before the initial
equilibration of the liquid phase, as the absence of polymer bonds facilitates the transition to the solid phase. The  pressure ensures similar  densities in the polymeric and 
atomic liquids.
After equilibration {for several relaxation times $\tau_{\alpha}$ in the liquid phase}, 
fifty-one runs spontaneously crystallized into two well defined classes. Seventeen runs formed solids quite close to face-centered cubic (Fcc) crystals 
and thirty-four runs formed Bcc-like atomic crystals. See sec. \ref{resultsdiscussion} for a detailed discussion. The remaining thirteen runs reached a variety of metastable solid-like conformations 
and were discarded.

\begin{figure}[t]
\begin{center}
\includegraphics[width=0.95\linewidth]{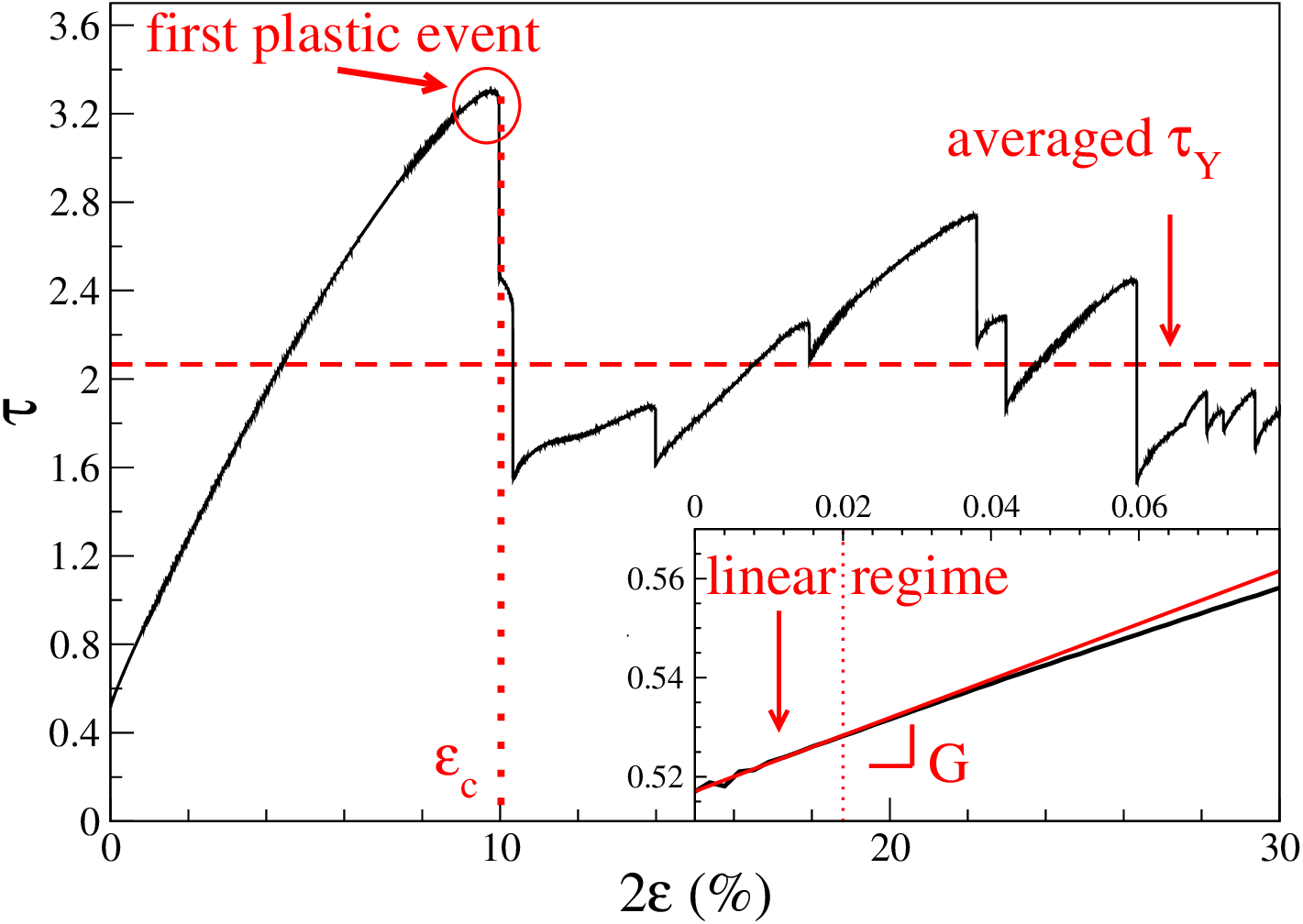}
\end{center}
\caption{Typical stress-strain curve under athermal quasi-static shear deformation of the semicrystalline polymer. After a first 'loading' phase, plastic events with macroscopic stress drops become apparent. $\tau_Y$ is defined as 
the average 
value of $\tau$ in the steady state phase \cite{LernerProcaccia}. $\epsilon_c$ is defined as the strain  at the first significant plastic event with stress drop of at least $\Delta\tau_{th}=0.1$. 
The elastic modulus $G$ (see inset) is measured via a linear fit of the stress-strain curve in the linear regime of small deformations
$2\epsilon<0.02$.}
\label{carico} 
\end{figure}

After completion of the solidification, all the systems were {quenched to temperature $T=10^{-3}$ and pressure $P=0$ in a time $\Delta t=0.003$ and, in agreement with others \cite{BarratShearReview}, later 
allowed to relax with an NPT run to let the total energy stabilize. The latter run lasted for a total time $\bar{t}=3000$.} The final densities of the polymeric and atomic Bcc-like solids are $\simeq 1.11$ and $\simeq 1.052$, respectively. The density offset is due to the different connectivity, having both solids the same pressure ( $P=0$) and temperature ( $T=0$ ). 

Simple shear deformations of the resulting athermal solids were performed via the Athermal Quasi-Static (AQS) protocol outlined in ref. \cite{BarratShearReview}. 
An infinitesimal strain increment $\Delta\varepsilon=10^{-5}$  is applied to a simulation box of side $L$ containing the sample, {after which the system is allowed to relax in the nearest 
local energy minimum with a steepest descent minimization algorithm. The accurate localization of the state corresponding to a local energy minimum ensures force equilibration on each particle, i.e. mechanical equilibration.}
The procedure is repeated until a total strain of $\Delta\varepsilon_{tot}=15 \cdot 10^{-2}$ is reached.
Simple shear is performed independently in the planes ($xy$, $xz$, $yz$), and at each strain step in the plane $\alpha\beta$ the corresponding component of the macroscopic stress tensor
$\tau_{\alpha, \beta}$ is taken as the average value of the per-monomer stress  $\tau_{\alpha, \beta}^i$: 
\begin{equation}
\label{stresstensor1}
\tau_{\alpha, \beta} =   \frac{1}{N}  \sum_{i=1}^N  \tau_{\alpha, \beta}^i
\end{equation}
In an athermal system the expression of the per-monomer stress in the atomic representation is \cite{allenMolPhys}: 
\begin{equation}
\label{stresstensor2}
\tau_{\alpha, \beta}^i =  \frac{1}{2 \, v} \sum_{j \ne i} r_{\alpha ij} F_{\beta ij} 
\end{equation}
where $F_{\gamma  k l}$ and $r_{\gamma  k l}$  are the $\gamma$ components of the force between the $k$th and the $l$th monomer and their  separation, respectively, and $v$ is the average per-monomer
volume, i.e. $v = L^3/N$.
For each plane a stress-strain curve is collected, an illustrative example of which is given  in Fig.\ref{carico}. 

Fig.\ref{carico} is quite analogous to what reported for many other systems under 
athermal conditions \cite{MottArgonSuter93,FalkViscoPlastic,Maeda2Dplastic,MelandroPlastic,LemaitrePlastic,ProcacciaPiecewise}  with an initial linear increase followed by increasing bending and onset 
of the plastic regime. In particular, similarly to other MD studies of glassy polymers \cite{YieldMDPolymerMM15}, one notices that, in the plastic regime,  the stress levels off to a plateau with 
fluctuations caused by subsequent loading phases and sudden stress drops. We point out that the initial non-zero stress in the unstrained solid seen in  Fig.\ref{carico} is a well-known phenomenon 
usually ascribed to the limited size of the simulation cell \cite{HawardCristPhysGlassyPolym}.

We measured the shear elastic modulus $G$ as the slope of the stress-strain curve in the linear regime,
within a strain threshold of $\varepsilon_{th}=0.01$, where the relation $\tau=2\epsilon\cdotp G$ holds, see Fig.\ref{carico} (inset).
Following Ref. \cite{LernerProcaccia}, the yield stress $\tau_Y$ is taken as the average value of the stress after the first significant plastic event, defined as the first stress drop 
of at least $\Delta\tau_{th}=0.1$ occurring at the critical strain $\epsilon_c$, see  Fig.\ref{carico}. This choice is  consistent with other definitions in the presence \cite{RottlerSoftMatter10}, or 
not \cite{RobbinsHoyJPolymSci06}, 
of strain softening, i.e. the reduction in stress following yield. The results are robust with respect to changes of $\Delta\tau_{th}$.

{
\begin{figure}[t]
\begin{center}
\includegraphics[width= \linewidth]{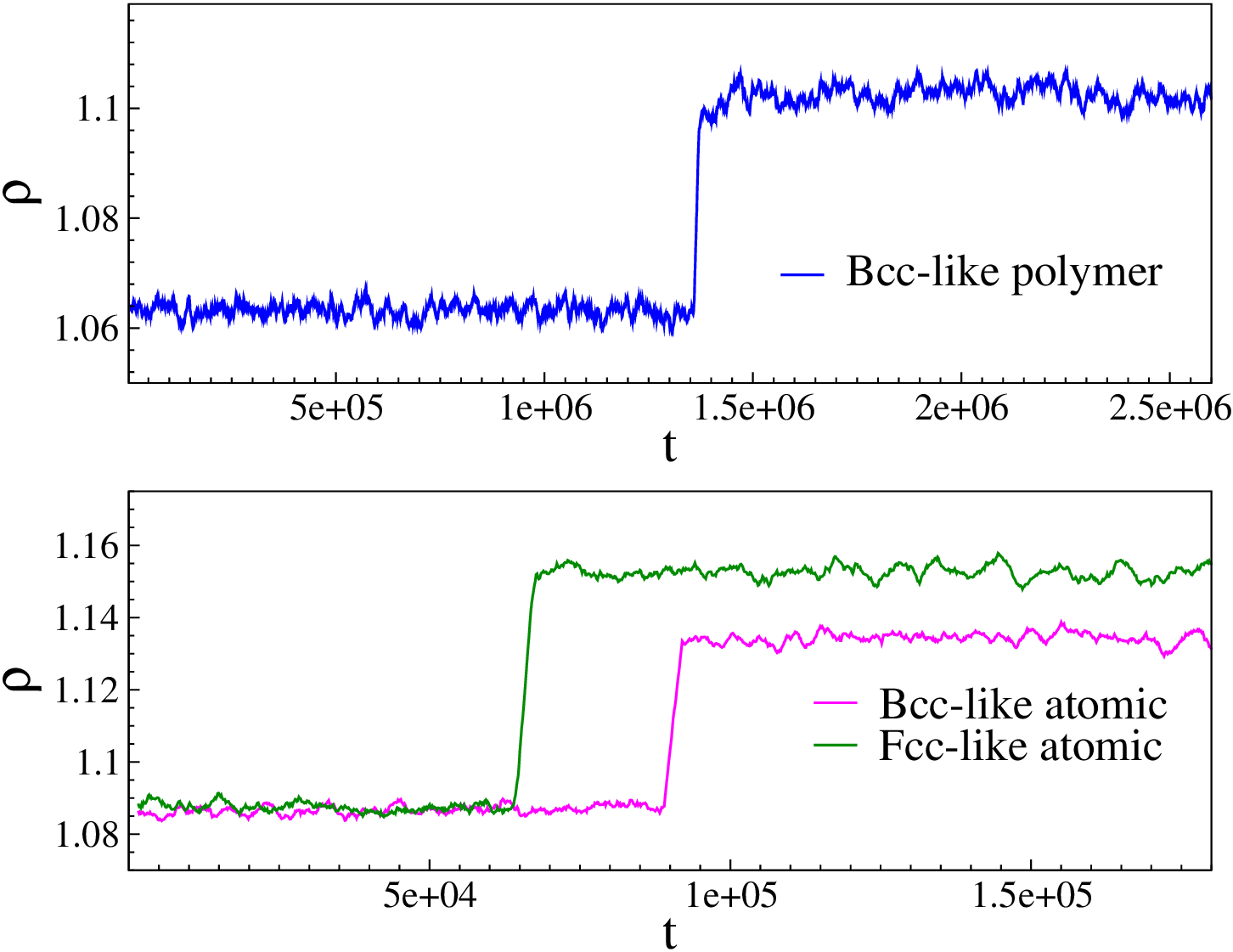}
\end{center}
\caption{Time dependence of the density of the systems under study. The selected lapses of time show the typical, abrupt jumps  for the polymer melt (top, $T=0.7$, $P=4.7$) and the atomic liquid 
(bottom, $T=1.5$, $P=20.0$)  signaling the spontaneous crystallization of liquid phases with very similar densities under isobaric, isothermal conditions. See text for details. Running averages were performed on the data to smoothen the noise.}
\label{VolDrop}
\end{figure}

\section{Results and discussion}
\label{resultsdiscussion}

\subsection{Solidification of the polymeric and the atomic liquids}
\label{spontaneous}

\noindent
Fig.\ref{VolDrop} plots typical runs during which solidification of the polymeric (top) and atomic (bottom) systems takes place. A single run is reported for each system under consideration. 
The crystallization is evidenced by the sudden increase of the density. Note that in the atomic liquids the size of the jump depends on the final crystalline state, as expected owing to the better 
packing of the Fcc lattice with respect to the Bcc one. The detailed characterization of the polymorphic structure of the polymer solid is reported elsewhere \cite{GiuntoliCristallo}. 
Notice that the jump is {\it smaller} for polymers ($\sim 3.3 \%$) than atomic liquids ($\gtrsim 4.5 \%$) even if the polymer melt has {\it lower} density. The finding agrees with the expectation
that high packing density is incompatible with connected structures \cite{ParkhouseKellyProcRoySoc95,GreavesNatMat2011}.

\subsection{Pre-shear structure of the athermal solids}
\label{structure}
We now characterize the structure of the athermal solids  {\it before} the shear deformation takes place.

\begin{figure}[t]
\begin{center}
\includegraphics[width= \linewidth]{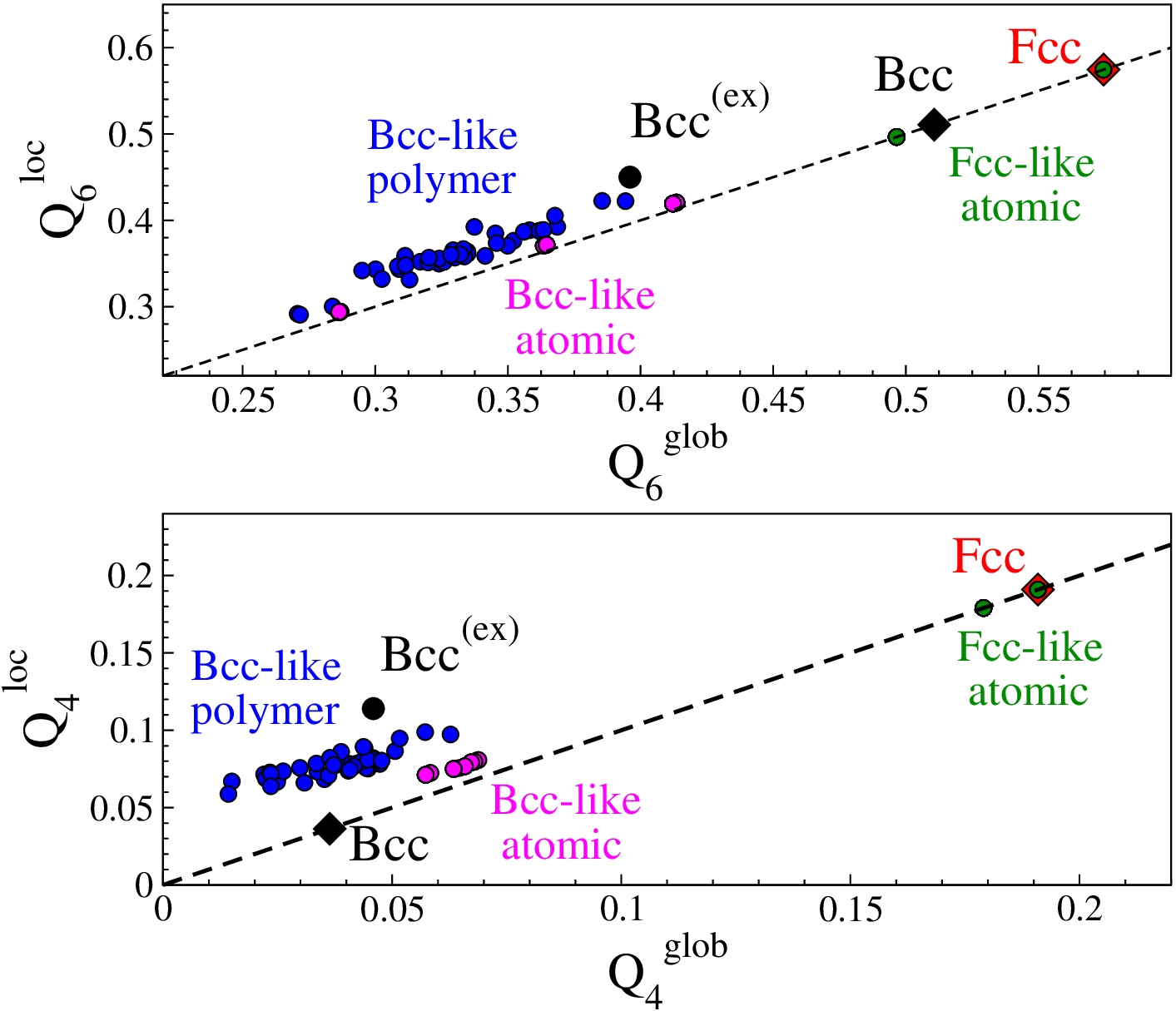}
\end{center}
\caption{Cross correlation $Q^{loc}_{l}$ vs $Q_l^{glob}$ with $l=6$ (top) and $l=4$ (bottom) of all the athermal solids under study, i.e. 42 polymer samples (blue dots) and 51 atomic samples (magenta 
{and green} dots).
The dashed line is the bisector $Q^{loc}_{l} = Q_l^{glob}$ corresponding to ideal order.
The pairs $(Q^{loc}_{l}, Q_l^{glob})$ with $l=4,6$ of the ideal Fcc and Bcc lattices ({red} and black diamonds) and the Bcc excited crystal (black dot) are also plotted. See text for details. 
The large size of the region enclosing the ($Q^{loc}_{l},Q_l^{glob})$ pairs for the polymer solid is ascribed to  significant polymorphism \cite{GiuntoliCristallo}.}
\label{coldOrder}
\end{figure}

\begin{figure}[t]
\begin{center}
\includegraphics[width= \linewidth]{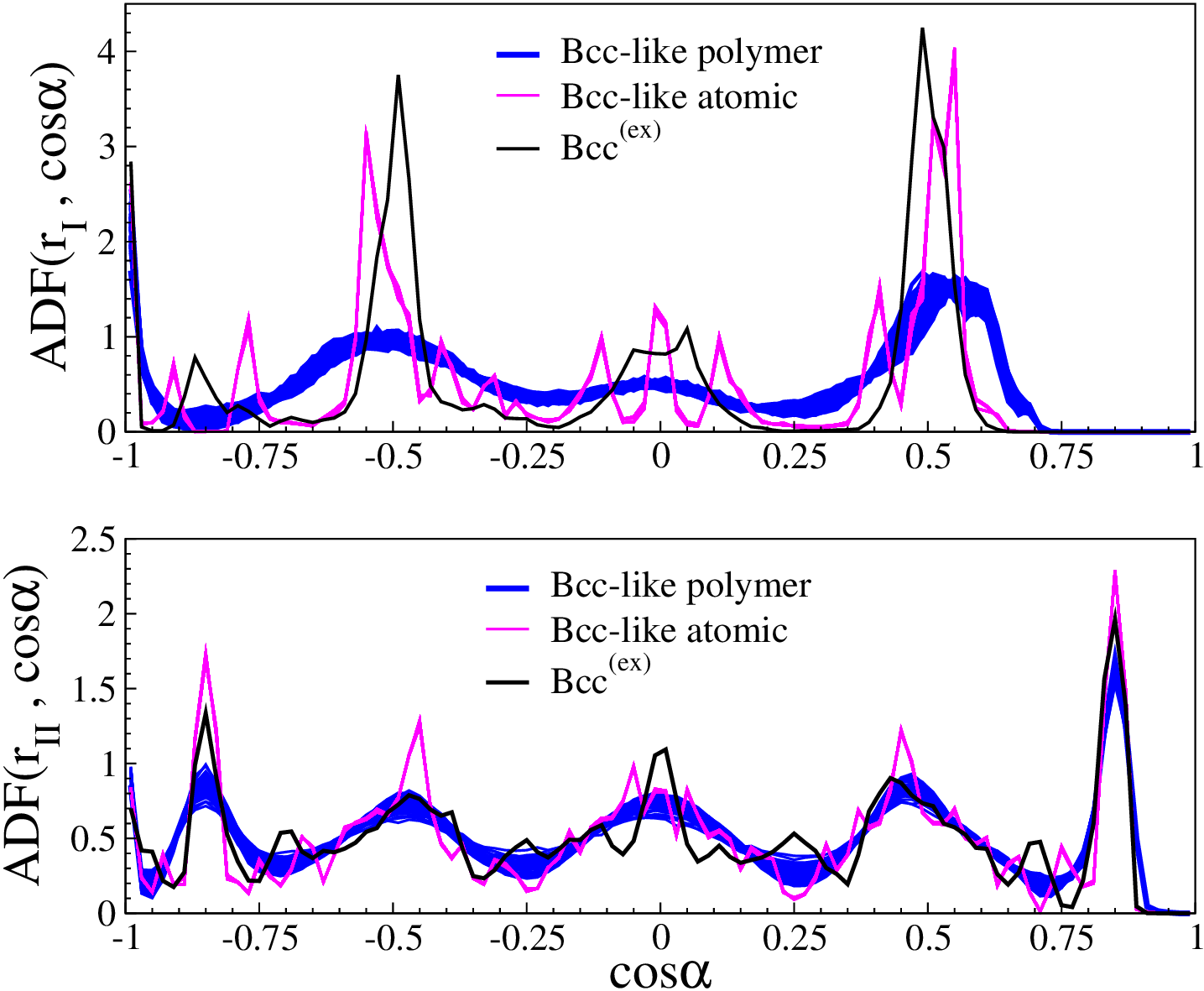}
\end{center}
\caption{ Angular distribution function (ADF) of the first (top) and the second (bottom) shells of all the athermal polymeric solids under study.}
\label{ADF}
\end{figure}

\begin{figure}[t]
\begin{center}
\includegraphics[width= \linewidth]{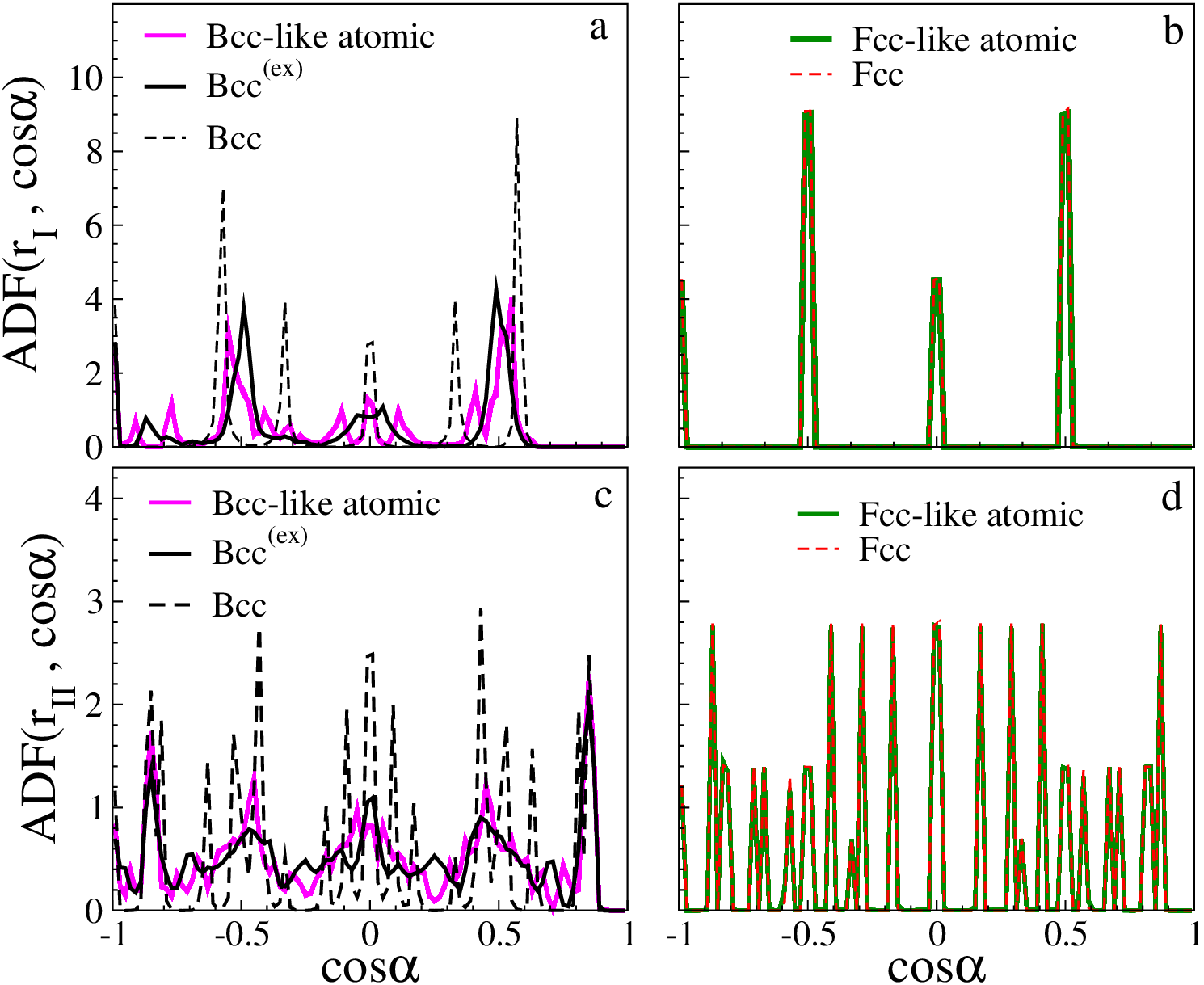}
\end{center}
\caption{Angular distribution function (ADF) of the first (panels: a, b) and the second (panels: c, d) shells of all the athermal atomic solids under study. Panels a,c and b,d refer to the 
Bcc-like and Fcc-like crystals, respectively.}
\label{ADFideali}
\end{figure}

\begin{figure}[t]
\begin{center}
\includegraphics[width=0.99\linewidth]{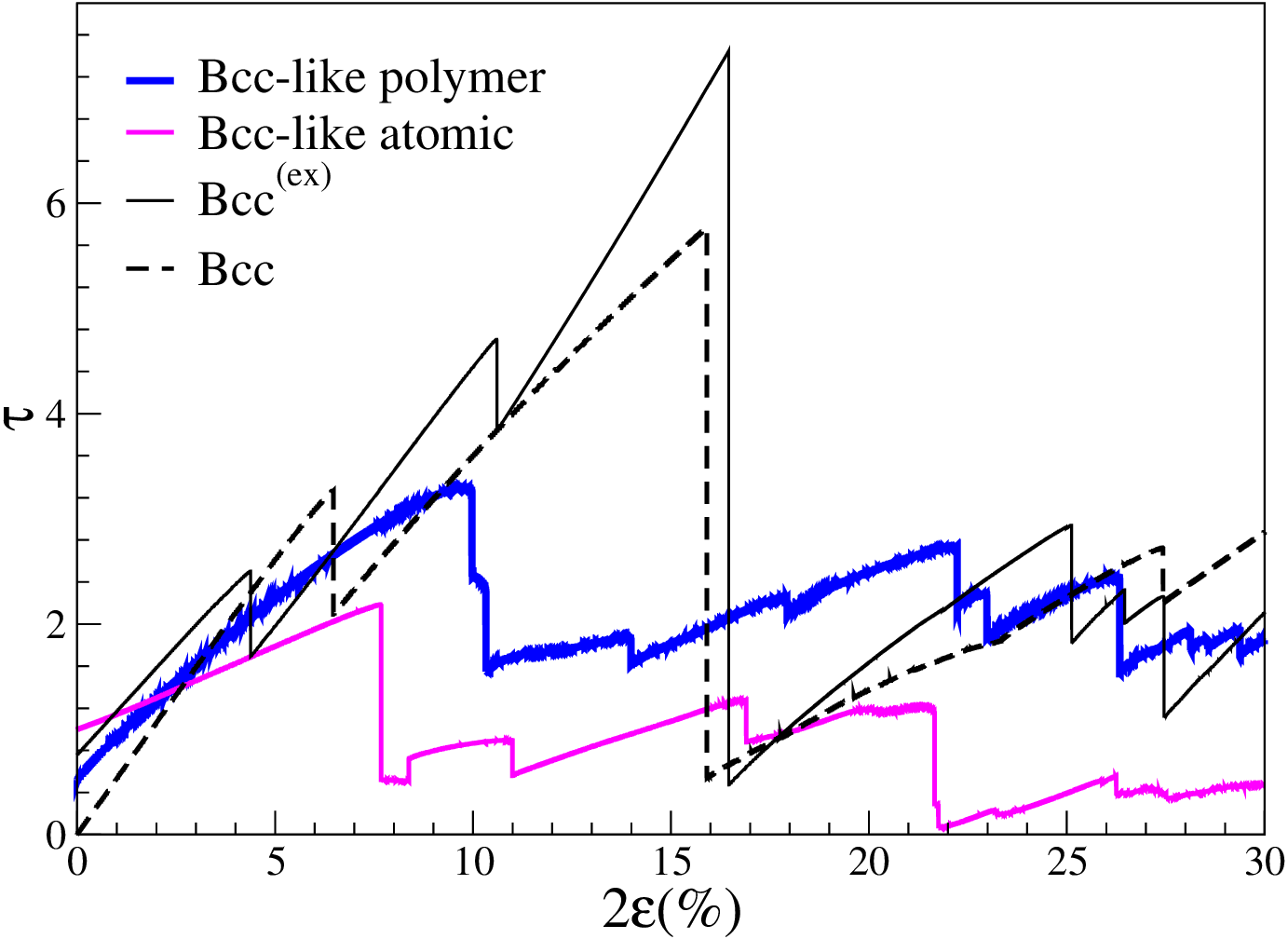}
\end{center}
\caption{Typical stress-strain curves during the quasi-static shear deformation of the athermal Bcc-like  polymeric and atomic solids. Curves pertaining to the ideal Bcc and the Bcc${}^{(ex)}$ crystals are also plotted. Notice that the  connectivity of the polymeric systems increases the number of abrupt changes of the stress in the plastic regime, resulting in a "noisy" pattern.}
\label{curves}
\end{figure}

To this aim, we compute the order parameters defined by Steinhardt \textit{et al.} \cite{Steinhardt83}.
One considers the polar and azimuthal angles $\theta({\bf r}_{ij})$ and $\phi({\bf r}_{ij})$ of the
vector ${\bf r}_{ij}$ joining the $i$-th central monomer with the $j$-th one belonging to the neighbors within a
preset cutoff distance $r_{cut} = 1.2 \; \sigma^* \simeq 1.35$ \cite{Steinhardt83}. $r_{cut}$ is a convenient definition of
the first coordination shell size \cite{sim}.
To define a global measure of the order in the system, one then introduces the quantity:
\begin{equation} \label{Qbarlm_global}
	\bar{Q}_{lm}^{glob}=\frac{1}{N_{b}}\sum_{i=1}^{N}
\sum_{j=1}^{n_b(i)}Y_{lm}\left[\theta({\bf
r}_{ij}),\phi({\bf r}_{ij})\right]
\end{equation}
where $n_b(i)$ is the number of bonds of $i$-th particle, $N$ is the total number of particles in the system, $Y_{lm}$
denotes a spherical harmonic and $N_b$ is the total number of bonds:
\begin{equation} \label{N_b}
	N_b=\sum_{i=1}^{N} n_b(i) 
\end{equation}
The global orientational order parameter $Q_l^{glob}$ is defined by:
\begin{equation} \label{Ql_glob}
 Q_l^{glob}=\left [ \frac{4\pi}{(2l+1)} \sum_{m=-l}^{l}
|\bar{Q}_{lm}^{glob}|^2 \right ]^{1/2}
\end{equation}
The above quantity is invariant under rotations of the coordinate system and takes characteristic values which can be used to quantify the kind and the degree of
rotational symmetry in the system \cite{Steinhardt83}. In the absence of {\it large}-scale order, the bond orientation is uniformly distributed around the unit sphere and $Q_l^{glob}$ 
is rather small \cite{RintoulTorquatoJCP96}. On the other hand, $Q_6^{glob}$ is very sensitive to any kind of crystallization and increases significantly when order appears \cite{GervoisGeometrical99}.
A local orientational parameter $Q_l^{loc}$ can also be defined. We define the auxiliary quantity
\begin{equation} \label{Qbarlm_local}
	\bar{Q}_{lm}^{loc}( i )=\frac{1}{n_b(i)}\sum_
{j=1}^{n_b(i)}Y_{lm}\left[\theta({\bf r}_{ij}),\phi({\bf
r}_{ij})\right]
\end{equation}
The local order parameter $Q_l^{loc}$ is defined as \cite{Steinhardt83}:
\begin{equation} \label{Ql_local}
 Q_l^{loc}=\frac{1}{N} \sum_{i=1}^{N}  \left [
\frac{4\pi}{(2l+1)} \sum_{m=-l}^{l} |\bar{Q}_{lm}^{loc}( i
)|^2 \right ]^{1/2}
\end{equation}
In general $Q^{loc}_{l}\ge Q_l^{glob}$. In the presence of ideal order, {\it all} the particles have the {\it same} neighborhood configuration, and the equality $Q^{loc}_{l} = Q_l^{glob}$
follows.

Cross correlations between $Q^{loc}_{l}$ and $Q_l^{glob}$ with $l=4,6$ proved to be rather useful to characterize the order of the solid phases \cite{GiuntoliCristallo}.
Fig.\ref{coldOrder} plots the cross-correlations for $l=6$ (top) and $l=4$ (bottom) for {all} the solids under study. To increase the readability, the plots also present the pairs 
$(Q^{loc}_{l}, Q_l^{glob})$ with $l=4,6$ corresponding to the ideal Bcc and Fcc atomic lattices. In addition, since the Bcc lattice is known  to be less stable then the 
Fcc one \cite{MisraCrystStability1940,MilsteinPRB70,GiuntoliCristallo}, the pairs $(Q^{loc}_{l}, Q_l^{glob})$ with $l=4,6$ of a Bcc excited crystal are also presented. 
The latter is obtained by heating the ideal Bcc crystal to temperature $T=0.7$ with $P \simeq 6.5$ and, after short equilibration, quenching it  at $T=10^{-3}$, $P=0$. 
The structure of the artificial excited atomic crystal, henceforth to be referred to as Bcc${}^{(ex)}$, was found to be nearly the same between $0.7 \le T \le 1.2$. The rationale behind the consideration of the Bcc${}^{(ex)}$ crystal is that the lack of stability of the ideal Bcc 
structure leads to significant deformations of the ordered structure obtained after the spontaneous crystallization \cite{GiuntoliCristallo}. 
In this respect, it seems more proper to compare 
our athermal solids with the athermal Bcc${}^{(ex)}$ solid rather than to the ideal Bcc one. The excitation of the Bcc lattice has been performed by using 8 different statistical configurations changing 
the velocities assigned to the particles to detect the possible presence of statistical differences between the runs. Such differences were not found in the structural analysis of the systems, but 
appear in elasticity and plasticity  (see sec. \ref{elastic}). Excitation was also tested on the Fcc lattice at the same temperature $T=0.7$, but no differences with the ideal structure were observed after the quench. 
The latter finding is consistent with the higher stability of the Fcc lattice with respect to the Bcc one \cite{MisraCrystStability1940,MilsteinPRB70,GiuntoliCristallo}.

Examination of Fig.\ref{coldOrder} leads to the following conclusions concerning the structure of the athermal solids before their deformation:

\begin{itemize}
\item solids are highly ordered since their characteristic points are close to the bisector;
\item atomic solids are either Fcc-like crystals or Bcc-like crystals. The former are quite close to the ideal structure whereas the latter, due to the lower stability of the Bcc lattice, 
exhibit some distribution and deviation from the ideality. The lower stability of the Bcc lattice is apparent in the well-separated locations of the points corresponding to the Bcc excited
crystal and the ideal Bcc crystal;
\item polymeric solids are: i) polymorphic, i.e. the corresponding blue dots are distributed, and  ii) exhibit Bcc-like structure, as evidenced by previous analysis  \cite{NicolaElastico}, and 
signaled by the localization of the dots close to the one of the Bcc excited crystal and the magenta dots of the Bcc-like atomic crystals.
\end{itemize}

Fig.\ref{coldOrder} provides some insight into the influence of the limited size of our sample on the local  and the global order of the Bcc-like polymeric polymorphs. To this aim, we compare the present MD results with previous ones \cite{NicolaElastico} concerning the same polymer model of interest here, bond length $\simeq 1.12 \,\sigma $ and number of monomers eight times larger than the present one. In the study of ref.\cite{NicolaElastico} 
crystallization occurs during quench-cooling since no nucleation was observed under isothermal condition, contrary to what reported in the present study. We found for the Bcc-like fraction $Q^{loc}_{4} \sim 0.10-0.15$ and $Q^{loc}_{6} \sim 0.38-0.42$ \cite{NicolaElastico}, to be compared to  $Q^{loc}_{4} \sim 0.05-0.1$ and  $Q^{loc}_{6} \sim 0.3-0.42$, see Fig.\ref{coldOrder}. This signals limited influence of the different  bond length, sample size and thermodynamic path to crystallization on the {\it local} order of the first coordination shell. As to the {\it global} order, the present result  $Q^{glob}_{6} \sim 0.27-0.42$ is quite {\it close} to the ideal value $Q^{loc}_{6}$, see Fig.\ref{coldOrder}, and somewhat {\it higher} than $Q_6^{glob} \sim 0.25$ of ref. \cite{NicolaElastico}. Tentatively, we ascribe the difference to the fact that both the isothermal crystallization and periodic boundary conditions favour better Bcc-like ordering in the present small sample than in a larger, quench-cooled sample.

\begin{figure}[t]
\begin{center}
\includegraphics[width=0.99\linewidth]{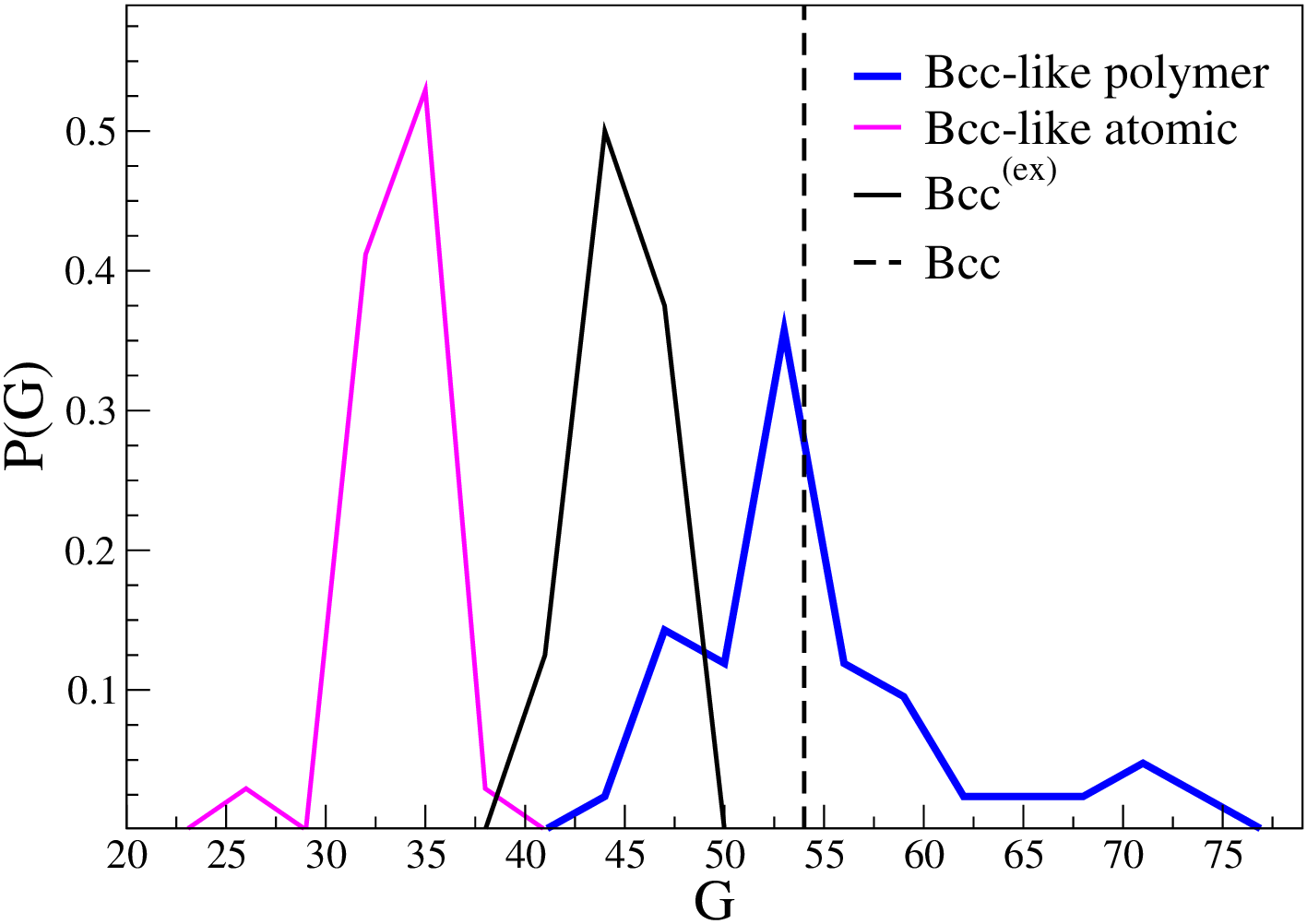}
\end{center}
\caption{The distributions of the shear elastic modulus for the athermal solids under study. It is seen that the elasticity of the polymer solid is distinctly {\it higher} than the ones of
the atomic solids with similar local structures (Bcc-like and Bcc${}^{(ex)}$). The width of the distribution for the polymer solid is due to the much larger polymorphism with respect to the atomic solids,
e.g. see Fig.\ref{coldOrder} and ref.  \cite{GiuntoliCristallo}. The softer character of the Bcc${}^{(ex)}$ solid with respect to the ideal Bcc crystal is apparent.}
\label{shearAve}
\end{figure}

Further insight into the local structure around the i-th particle is offered by the angular distribution function $ADF( \cos \alpha_{jk})$  where $\alpha_{jk}$ is the angle between ${\bf r}_{ij}$ and ${\bf r}_{ik}$, and the vector ${\bf r}_{ij}$ joins 
the $i$-th central particle with the $j$-th one which is $r_{ij}$ apart. The ADF analysis is carried out on the first and the second neighbor shells surrounding the i-th particle. 
They are singled out by the constraints $r_{min} \le r_{ij}, r_{ik} \le r_{max}$ with  $r_{min}=0.8, r_{max}=1.35$ (first shell) and $r_{min}=1.35, r_{max}=2.2$ (second shell) \cite{GiuntoliCristallo}. 
{Note that the "first shell" considered by the ADF analysis is virtually the same region considered by the Steinhardt order parameters 
( $r <r_{cut} \simeq 1.35$) since the number of monomers spaced by less than $r_{min}=0.8$ is negligible. }

Fig.\ref{ADF}  shows the ADF of all the athermal polymeric solids under study. The differences between the polymeric ADF and the atomic (Bcc-like and Bcc${}^{(ex)}$) 
ADFs in the first shell are ascribed to the fact that the bond length of the polymeric chain is incommensurate with the atomic lattice \cite{NicolaElastico}. 
The connectivity effect is negligible in the second shell and the deviations are quite smaller. For clarity reasons, the ADF of the ideal Bcc lattice is not shown due to the rather distinct pattern, 
see {Fig.}\ref{ADFideali}.
{The ADF analysis in Fig.\ref{ADF} clarifies that the agreement between the particle arrangements 
of the polymeric and the atomic athermal solids is partial in the first shell (sensed by the Steinhardt parameters) but  rather good in the second shell}.

Fig.\ref{ADFideali} plots the ADF of all the athermal atomic solids under study. It is seen the ADF of the Bcc-like fraction is well accounted for by the ADF of the Bcc${}^{(ex)}$ 
lattice {both in} the first and the second shell whereas the deviations of the ideal Bcc crystal are large. Instead, the ADF of the Fcc-like fraction is rather close to the ADF of the ideal 
Fcc lattice {both in} the first and the second shell.

\subsection{Elastic and plastic response}
\label{elastic}

The elastic and plastic response of the polymeric and the atomic solids are now examined. We focus on systems with similar, i.e. Bcc-like and and Bcc${}^{(ex)}$, local environment. 
Related, illustrative Stress-Strain curves are given in Fig.\ref{curves}. The complete sets of curves for all the systems under study  are used to draw the elastic modulus $G$, the critical strain
$\epsilon_c$ of the first plastic event and the average yield stress $\tau_Y$. Suitable averages over the three \textit{xy, xz, yz} deformation planes are taken for each run.

Fig.\ref{shearAve} plots the distributions of the elastic modulus $G$ of the polymeric and the atomic systems with rather similar local structure. It is seen that the polymeric system has {\it larger} 
shear modulus. Notice that the comparison must be performed with the physical Bcc-like atomic solid and {\it not} the artificial Bcc${}^{(ex)}$ one which is presented for reference only. The elastic modulus of the ideal Bcc crystal is indicated to show the softening effect of the preparation of the Bcc${}^{(ex)}$ solid.

Fig.\ref{stress} plots the distributions of the shear strength of the polymeric and the atomic systems with rather similar local structure. It is seen that the polymeric system has {\it larger} 
strength than the atomic Bcc-like solid. The strength is comparable to the one of the artificial Bcc${}^{(ex)}$ atomic solid and the ideal Bcc crystal. 

\begin{figure}[t]
\begin{center}
\includegraphics[width=0.99\linewidth]{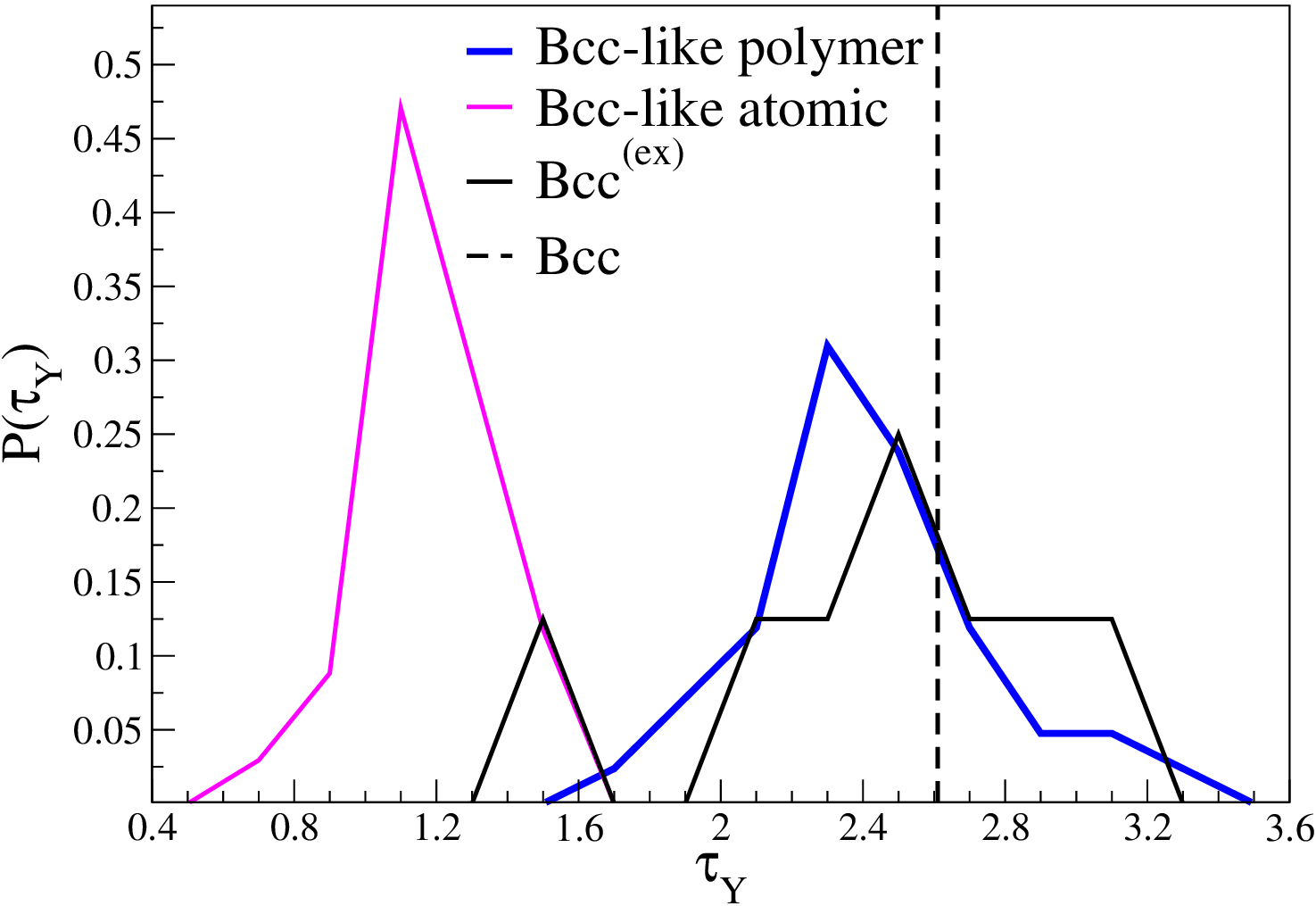}
\end{center}
\caption{The distributions of the yield strength for the athermal solids under study. It is seen that the strength of the polymer solid is distinctly {\it higher} than the one of the atomic 
solid with similar local structure (Bcc-like) and comparable to the ones of the artificial atomic solid Bcc${}^{(ex)}$ and the ideal Bcc crystal.}
\label{stress}
\end{figure}

\begin{figure}[t]
\begin{center}
\includegraphics[width=0.99\linewidth]{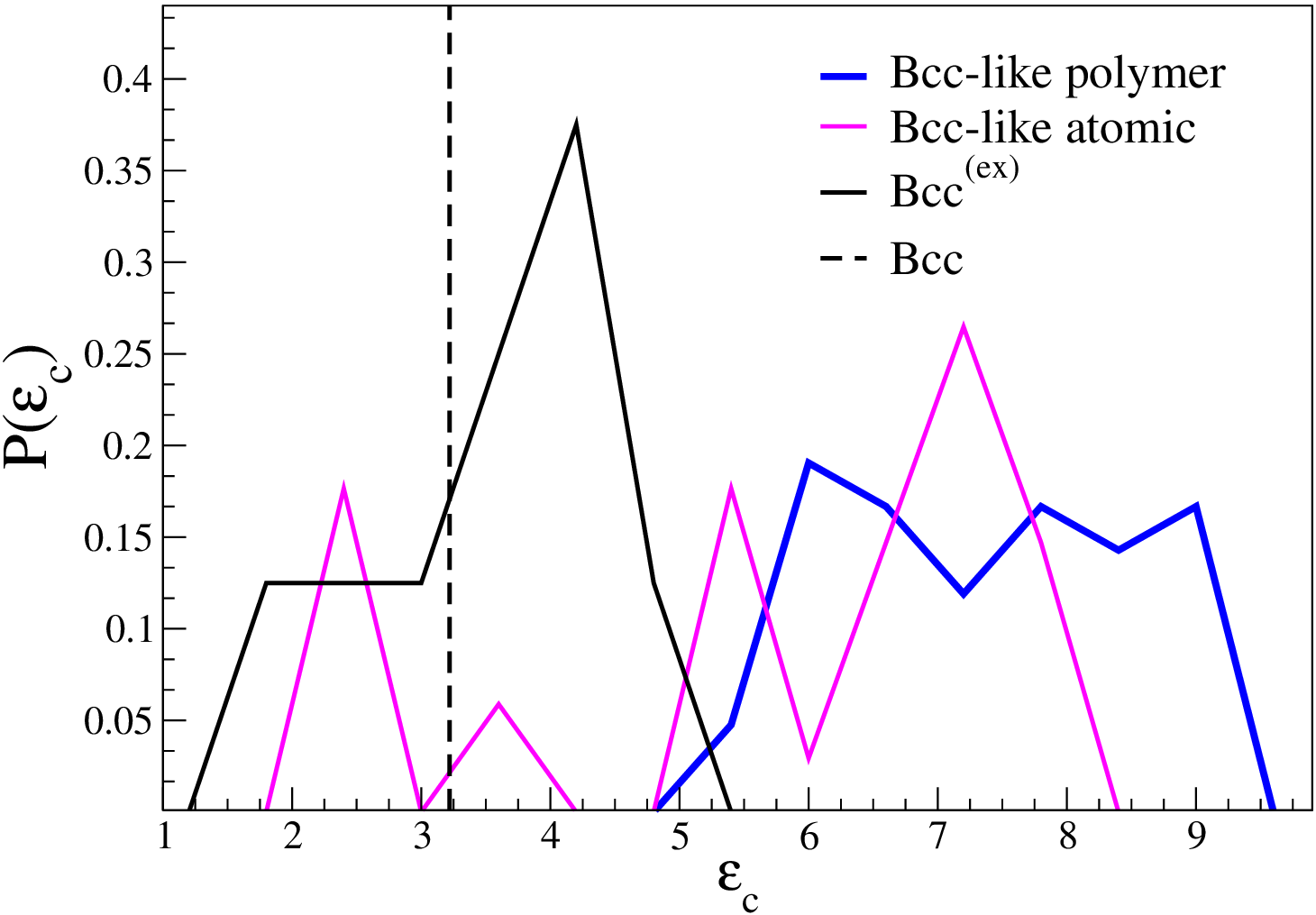}
\end{center}
\caption{The distributions of the critical strain $\epsilon_c$ for the athermal solids under study. It is seen that the strain of the polymer solid is comparable to the one of the atomic 
solid with similar local structure (Bcc-like) and distinctly {\it higher} than the one of the artificial atomic solid Bcc${}^{(ex)}$  and the ideal Bcc crystal.}
\label{strain}
\end{figure}

Finally, Fig.\ref{strain} shows the critical strain $\epsilon_c$ at which the first plastic event is observed. It is seen that the strain of the polymer solid is comparable 
to the one of the atomic solid with similar local structure (Bcc-like) and distinctly {\it higher} than the one of the artificial atomic solid Bcc${}^{(ex)}$ and the ideal Bcc crystal.

The above findings suggest that the disordered connectivity of the chains suppresses dislocation-mediated plasticity  in polymeric solids with positional order.

As a final remark, we point out that the increases of the modulus ( $\sim 60\%$) 
and the strength ( $ \sim 100 \%$ ) of the  Bcc-like polymeric athermal solid with respect to the corresponding atomic one cannot be ascribed to the slightly larger density of the 
former with respect to the latter ( $ \sim 5.5 \%$ ). 
In fact, a density increase from $1.04$ to $1.15$ ( $\sim 10 \%$) caused by spanning different bond lengths from $1.03$ to $0.91$ has been proved to minimally rise the elastic modulus ($\sim 9\%$)
and the yield stress ($\sim 13\%$) of polymeric solids at $T=0$, $P=0$ \cite{NicolaElastico}.

\section{Conclusions}
\label{conclusions}
We perform thorough MD simulations to compare elasticity and yielding of atomic crystals and model semicrystalline polymers with fully-flexible chains (no bending potential). 
Both the atomic and the polymeric solids have very similar, Bcc-like {\it positional ordering} of the particles. 
We find that the elastic modulus and the shear yield strength are higher in semicrystalline polymers with respect to atomic crystals, 
whereas the critical yield strain $\epsilon_c$ are comparable. The findings suggest that the disordered connectivity of the chains suppresses dislocation-mediated plasticity 
in polymeric solids with positional order.

\section*{Acknowledgements}
\label{Ack}
We thank Francesco Puosi{, Sebastiano Bernini} and Nicola Calonaci for helpful discussions. 
A generous grant of computing time from IT Center, University of Pisa and Dell${}^\circledR$ Italia is gratefully acknowledged.

\bibliography{biblio.bib}

\end{document}